\documentclass[3p,twocolumn]{elsarticle}

\usepackage{graphicx}
\usepackage{amssymb}
\usepackage{xcolor}
\usepackage{dcolumn}
\usepackage{ulem}
\usepackage[utf8]{inputenc}
\usepackage[LSF,LGR,T3,T1]{fontenc}
\newcolumntype{d}[1]{D{.}{.}{#1}}

\DeclareUnicodeCharacter{2010}{–}
\DeclareUnicodeCharacter{2212}{—}

\usepackage{booktabs} 
\usepackage[version=3]{mhchem}
\usepackage{textcomp}
\usepackage[squaren]{SIunits}
\usepackage{xspace}
\usepackage[colorinlistoftodos]{todonotes}
\newcommand \wn {cm$^{-1}$\xspace}

\graphicspath{{./Figs/}}
\newcommand \ub {\textmu bar\xspace} 

\newcommand \esi {Supporting Information}

\newcommand{\rev}[1]{\textcolor{blue}{#1}}

\newcommand{\phc}{\ce{PhC3N}\xspace}

\graphicspath{{./Figs/}}

\journal{Journal of Molecular Spectroscopy}

\begin{document}

\begin{frontmatter}


\title{A rotational and vibrational investigation of phenylpropiolonitrile (\ce{C6H5C3N})}



\author[label1,label2]{Zachary Buchanan\corref{cor1}}
\cortext[cor1]{zsbuchanan@ucdavis.edu}
\author[label3]{Kin Long Kelvin Lee}
\author[label1]{Olivia Chitarra}
\author[label3]{Michael C. McCarthy}
\author[label1,label4]{Olivier Pirali}
\author[label1]{Marie-Aline Martin-Drumel}

\address[label1]{Universit\'e Paris-Saclay, CNRS, Institut des Sciences Mol\'eculaires d'Orsay, 91405 Orsay, France}
\address[label2]{Department of Chemistry, The University of California Davis, Davis, CA, USA}
\address[label3]{Center for Astrophysics $\vert$ Harvard \& Smithsonian, Cambridge, Massachusetts 02138, United States}
\address[label4]{SOLEIL Synchrotron, AILES beamline, l'Orme des Merisiers, Saint-Aubin, 91190 Gif-sur-Yvette, France}

\begin{abstract}

{
The evidence for benzonitrile (\ce{C6H5CN}) in the starless cloud core TMC--1 makes high-resolution studies of other aromatic nitriles and their ring-chain derivatives especially timely. 
One such species is phenylpropiolonitrile (3-phenyl-2-propynenitrile, \ce{C6H5C3N}), whose  spectroscopic characterization is reported here for the first time.
The low resolution (0.5 \wn) vibrational spectrum of \ce{C6H5C3N} has been recorded at far- and mid-infrared wavelengths (50--3500 \wn) using a Fourier Transform interferometer, allowing for the assignment of band centers of 14 fundamental vibrational bands. 
The pure rotational spectrum of the species has been investigated using a chirped-pulse Fourier transform microwave (FTMW) spectrometer (6--18 GHz), a cavity enhanced FTMW instrument (6--20 GHz), and a millimeter-wave one (75--100 GHz, 140--214 GHz).
Through the assignment of more than 6200 lines, accurate ground state spectroscopic constants (rotational, centrifugal distortion up to octics, and nuclear quadrupole hyperfine constants) have been derived from our measurements, with a plausible prediction of the weaker bands through calculations.
Interstellar searches for this highly polar species can now be undertaken with confidence since the astronomically most interesting radio lines have either been measured or can be calculated to very high accuracy below 300\,GHz.}
\end{abstract}

\begin{keyword}
pure rotation \sep vibration \sep astrophysical species \sep PAH derivative \sep phenylpropiolonitrile
\end{keyword}

\end{frontmatter}


\section{Introduction}

The ubiquity of aromatic molecules is closely-correlated to their stability and lack of reactivity, with functionalized aromatics serving as a common motif in biological chemistry.  Polycyclic aromatic hydrocarbons are a prominent class of aromatics; they are well-known constituents in the outflows of certain evolved carbon stars and common byproducts in incomplete combustion processes \cite{frenklach_formation_1989,snow_interstellar_1995,tielens_interstellar_2008}. 
%
From a purely spectroscopic viewpoint, a number of simple derivatives of benzene ---the prototypical aromatic ring \ce{C6H6}--- have either not been characterized at all or at insufficient resolution to undertake an astronomical search in the coldest most quiescent molecular clouds in space.
The recent discovery of benzonitrile ---the simplest aromatic nitrile (\ce{C6H5CN} or \ce{PhCN} hereafter)--- using radio observations towards Taurus Molecular Cloud (TMC-1) \cite{mcguire_detection_2018} has reignited the interest in nitrogen-containing aromatics generally and CN-functionalized aromatics specifically \cite{McNaughton2018}. The subsequent identification of cyanocyclopentadiene, \ce{C5H5CN}, in the same cloud \cite{mccarthy_interstellar_2020} has only intensified this interest.
Nitriles are also known to be important constituents in the chemistry of Titan's atmosphere \cite{gautier:hal-00756910}, and they are very prominent in the interstellar medium (ISM), accounting for roughly 20\% of the 220 or so molecules (47) detected in the ISM to date, including cyanopolyynes as long as \ce{HC11N} \cite{Loomis2020}.

The presence of a nitrile group often imparts a molecule with a large permanent dipole moment and an intense rotational spectrum.  In the case of \ce{PhCN}, replacing a single H atom in benzene with a CN group transforms an otherwise highly symmetric  ring into a highly polar species ($\mu_a = 4.5$\,D; \cite{wohlfart_precise_2008}) thereby greatly aiding its detection both in the laboratory and in space.
Whilst the rotational spectrum of \ce{PhCN} has been known for more than half-a-century \cite{bak_benzene_1962}, its interstellar detection was greatly aided by measurements at very high accuracy ($i.e.$,~at a resolving power $f/\Delta f >10^6$) at centimeter wavelengths \cite{mcguire_detection_2018,wohlfart_precise_2008}.
In light of this finding,  new ---or in many cases improved--- high-resolution studies of molecules closely related in structure and composition are worth pursuing. 

\begin{figure}[ht!]
    \centering
    \includegraphics[width=0.45\textwidth]{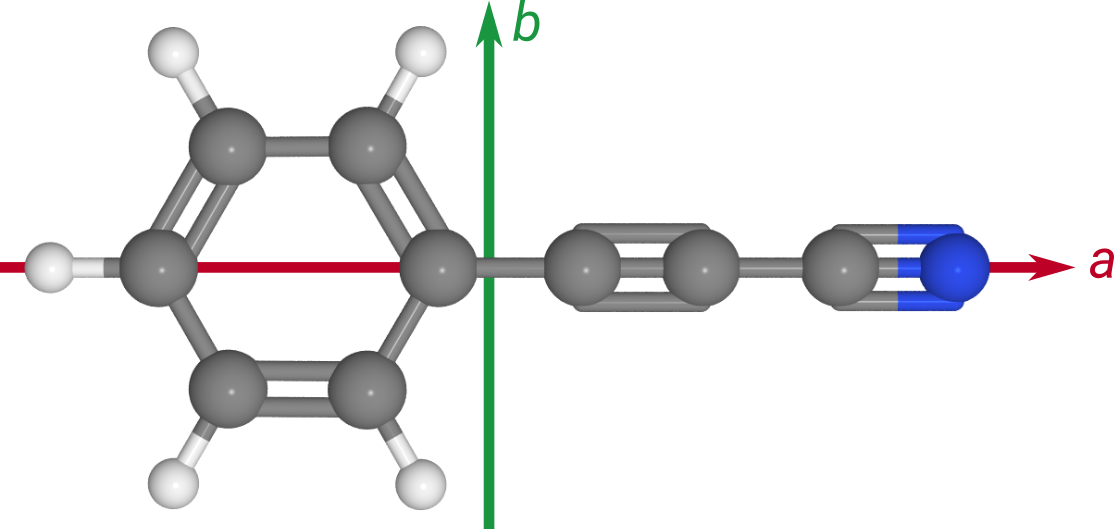}
    \caption{Molecular structure of \phc; $a$ and $b$ principal inertial axes are indicated (in red and green, respectively); the $c$-axis extends out of the molecular plane and is not shown.}
    \label{fig:phc3n}
\end{figure}

In this work, we report a combined pure rotational and vibrational investigation of a derivative of \ce{PhCN}, phenylpropiolonitrile (3-phenyl-2-propynenitrile, \ce{C6H5C3N}, abbreviated as \ce{PhC3N} in the following) where the nitrile group is replaced by a longer chain variant (\ce{C3N})  (Fig.~\ref{fig:phc3n}). 
This species has previously been identified as a possible product from the reaction between the cyano radical and  phenylacetylene (\ce{C6H5CCH}) \cite{bennett_chemical_2010}. To the best of our knowledge, however, spectroscopic investigations have been limited to experimental and theoretical vibronic spectroscopy \cite{Jawad2017,Viquez2017} while high-resolution, rotationally-resolved studies are apparently lacking.  In light of the large permanent electric dipole moment calculated here (5.9\,D), laboratory measurement of rotational frequencies would allow astronomical searches for this ring-chain to be undertaken with little or no ambiguity.  If found in space, the abundance of \phc would provide a key test for models of aromatic chemistry which are poorly constrained at present.   
The infrared spectrum of \ce{PhC3N} is also of interest as a point of comparison with other benzene derivatives whose vibrational spectra are often plagued by {a myriad of perturbations and resonances, notably Fermi and Darling-Dennison in the \ce{C-H} and \ce{C#C} stretching regions \cite{king_ethynylbenzene_1970,chang_vibronic_2010,pirali_high-resolution_2009}, in addition to Fermi and Coriolis interactions for low frequency (${\sim}150$\,\wn), large amplitude modes that are prominent at room temperature \cite{pirali_rotation-vibration_2015,kisiel_millimetre-wave_2010,zdanovskaia_103360_2018}---even under astronomical conditions.}

\section{Experimental and computational methods}\label{sec:method}
    \subsection{Quantum chemical calculations}
    Calculations were performed using the Gaussian'16 suite of electronic structure programs \cite{frisch_gaussian_2016}. The goal of these calculations was to provide accurate estimates of spectroscopic parameters, including rotational constants, dipole moment, rotation-vibration corrections, and fundamental vibrational frequencies using second-order vibrational perturbation theory (VPT2). 
    Geometry optimizations of \ce{PhC3N} and four of its isomers---namely the isocyanide analog \ce{PhCCNC} and three variants where \ce{CN} is substituted for a H atom on the ring of phenyl acetylene to yield \textit{ortho-, meta-,} and \textit{para}-cyanoethynylbenzene (CEB)---were carried out at the $\omega$B97X-D/cc-pVQZ level on an ultrafine integration grid, in which optimized structures are those in which convergence to better than $10^{-5}$ of the root-mean-squared (RMS) value of the gradient has been achieved. 
    For \ce{PhC3N}, we also performed (an)harmonic frequency analysis, obtaining both the harmonic and anharmonic vibrational frequencies and intensities at the same level of theory; with the latter, the rotation-vibration interaction constants $\alpha$ were  computed. 
    Cartesian coordinates of the optimized equilibrium structures can be found in Tables S1-S5 in the \esi.

    In addition to spectroscopic parameters, we have also performed rudimentary thermochemical calculations on the relative energies of \ce{PhC3N} and its isomers using the G3//B3LYP composite method \cite{baboul_gaussian-3_1999}, which has been shown to provide near-chemical accuracy at excellent computational cost \cite{simmie_benchmarking_2015}. 
    Given that semi-empirical methods typically perform best on closed-shell molecules with relatively simpler electronic structure, as the ones studied here, we believe these calculations provide a quantitative determination of relative stabilities, accurate to $\pm4$\,kJ/mol.

    \subsection{Fourier-transform infrared measurements}
    The gas-phase vibrational spectrum of \phc was recorded in the far-infrared (far-IR) and mid-infrared (mid-IR) using the Bruker IFS 125 FT interferometer located at the AILES beamline of the SOLEIL synchrotron facility (no synchrotron radiation was used in the present study) \citep{Brubach2010}.  
    For the far-IR measurements, the spectrometer was equipped with a $6$\,{\textmu}m mylar-silicon composite beamsplitter and a liquid helium-cooled silicon bolometer. 
    A KBr beamsplitter and a sensitive HgCdTe detector, equipped with a cryogenically cooled entrance iris and optical filters \citep{Faye2016}, were used in the mid-IR region.
    Vapor of \ce{PhC3N} was injected in a White-type multipass cell aligned to yield a 150\,m optical path length \citep{Pirali2013} and isolated from the interferometer by 50\,{\textmu}m-thick polypropylene windows in the far-IR and wedged ZnSe windows in the mid-IR range. 
    In both spectral regions, the interferometer was continuously evacuated to $10^{-5}$\,mbar to minimize absorption from residual water.
    Spectra were recorded at a resolution of 0.5\,\wn using a globar light source and an entrance iris of 4\,mm, and consist of 100 and 500 co-added interferograms for the far-IR and mid-IR regions, respectively. 
    Both spectra were recorded at a sample pressure of 5\,\ub.
    It is worth noting that the rotational structure within the vibrational bands could not be resolved even at the highest resolution of the spectrometer (0.001\,\wn).

    \subsection{Chirped-pulse Fourier-transform microwave measurements}
    Microwave measurements were performed using a chirped-pulse Fourier transform microwave (CP-FTMW) spectrometer  located at the Center for Astrophysics \cite{Crabtree2016} which operates between 8 and 18\,GHz.
    About 0.3\,g of solid \phc was introduced into a reservoir located behind the pulsed nozzle and Ne carrier gas (at a flow of $\sim$20\,sccm at standard temperature and pressure) passed through the reservoir; the resulting gas mixture was then injected into the vacuum chamber by operating the pulsed valve at a very low repetition rate  (5\,Hz).
    Because the vapor pressure of \phc is relatively high at room temperature (several mbar), it was not necessary to heat the reservoir  to observe rotational lines in the CP-FTMW spectrum with good signal-to-noise ratios. 
    
\begin{figure}[ht!]
        \centering
        \includegraphics[width=\columnwidth]{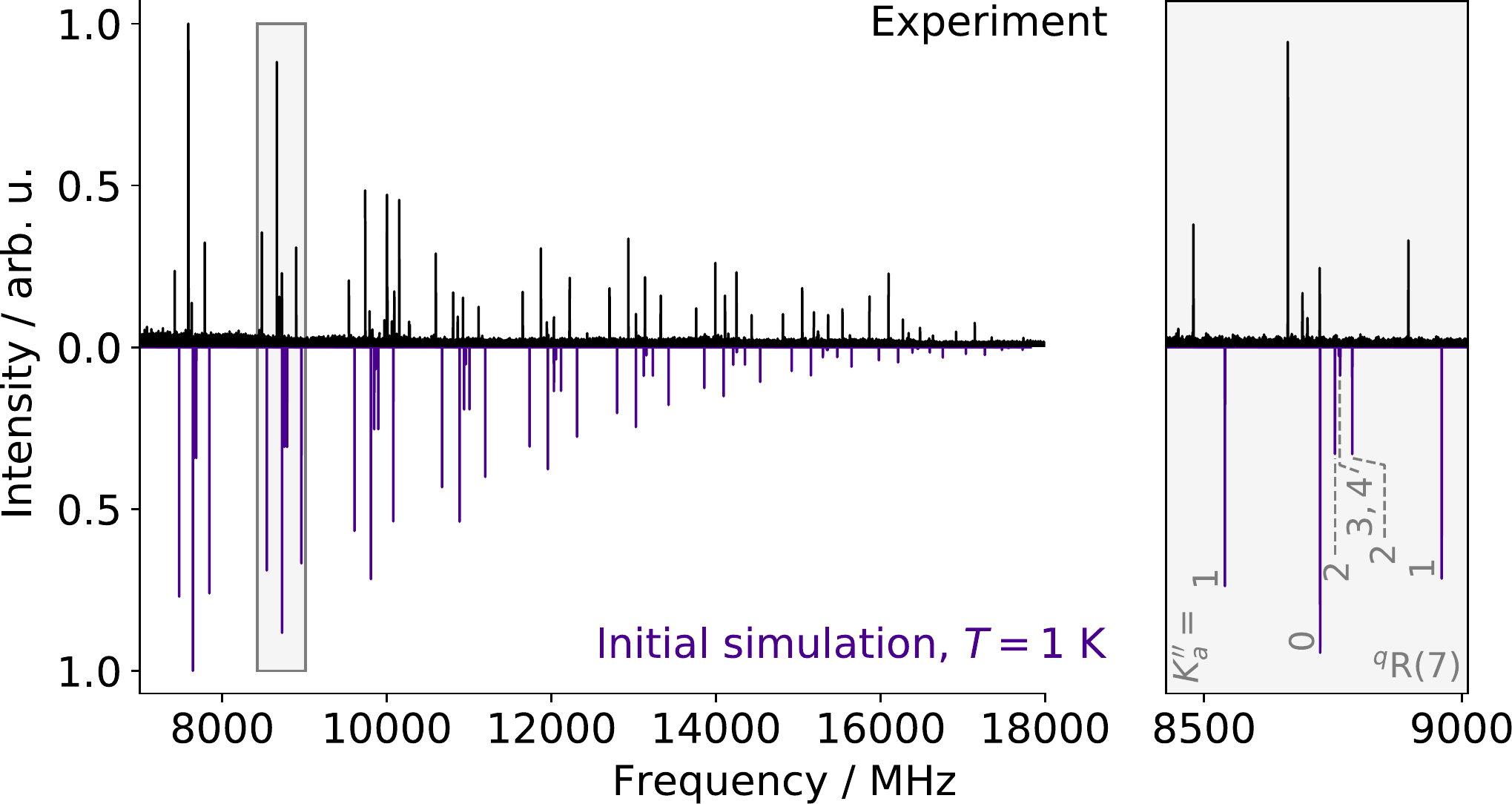}
        \caption{Experimental CP-FTMW spectrum (in black) of \ce{PhC3N} in comparison with a simulation at $T_\mathrm{rot}=1$\,K  (in purple) using the calculated set of rotational constants; the simulation is plotted here with negative intensity to more easily compare the two spectra. \textit{Left panel:} the full spectrum, noting that a few lines originating from impurities are visible around 10\,GHz; \textit{Right panel:} expanded view around the $^\mathrm{q}$R(7) transitions, as  indicated by the gray rectangle on the left panel. The simulation was performed using PGOPHER \cite{Pgopher}.}
        \label{fig:CPMW}
    \end{figure}

    Approximately 17,000 nozzle pulses, each probed by 10 microwave chirps, were acquired.
    Additional details on the experimental set-up are provided in Ref.~\cite{Crabtree2016}.
    The resulting spectrum, with electronics artifacts and well-known contaminant lines ($e.g.$, acetone) removed, is displayed in Fig.~\ref{fig:CPMW}.
    Transitions with $5 \leq J''\leq 15$ and $K_a'' \leq 4$ are visible.
    
    
    \subsection{Cavity-based Fourier-transform microwave measurements}
    In parallel to the present investigation, pure rotational  transitions  of  \phc  were  also  identified by several of the co-authors of this paper while analyzing the  discharge products of a benzene/nitrogen (\ce{N2}) mixture \cite{McCarthy2020} . 
    Using a cavity-enhanced FTMW (CE-FTMW) spectrometer operating between 6 and 40\,GHz, transitions of \phc were measured at roughly ten times  higher spectral resolution than can be achieved with the CP-FTMW instrument. In the following, we only briefly summarize the experimental conditions relevant to this work; further details about the CE-FTMW instrument and the discharge mixture experiments are presented in Refs. \cite{Crabtree2016} and \citep{McCarthy2020}, respectively. 
    
    \phc was synthesized by subjecting a mixture of \ce{C6H6} and \ce{N2} diluted heavily in \ce{Ne} to a high-voltage discharge (800\,V); typical flow rates were 14, 12, and 20\,sccm respectively. Compared to the CP-FTMW measurements where \phc was the precursor, the discharge experiment results in a slightly higher rotational temperature, $i.e.$, $\sim 10$\,K and produces rotational lines with demonstratively lower signal-to-noise ratio.
    Nevertheless, at the highest resolving power ($\sim10^7$) the nitrogen nuclear-quadrupole structure for several low-$K_a$ transitions ($K_a^{\prime\prime}=0-2$) in the 6--20\,GHz region was partially resolved.

    \subsection{Absorption millimeter-wave measurements}
    The room temperature spectrum of \phc was recorded at ISMO with an absorption spectrometer in which a frequency multiplication scheme is used to generate millimeter-wave radiation   \cite{Pirali2017}.
    By combining the output of a radiofrequency synthesizer (Rohde \& Schwarz) operating in the 2--20\,GHz region with one of two amplifier / multiplier chains, it is possible to produce broadly tunable radiation with modest power (of a few mW) throughout the millimeter band; a Radiometer Physics GmbH (RPG) SMZ110 for 75--97\,GHz, and a Virginia Diodes Inc.~(VDI) for 141--214\,GHz. 
    
    The millimeter-wave radiation was collimated using a 10\,mm focal length Teflon lens into a 2\,m long Pyrex absorption cell and further focused onto a Schottky diode detector from VDI,  using a second identical lens. The input radiation was frequency modulated at a frequency of 48.2\,kHz and a commercial lock-in amplifier (Ametek)  demodulated the signal at the second harmonic.   The spectrum was recorded using 50\,kHz frequency steps, a time constant of 50/100\,ms, and an FM deviation of 200/250\,kHz (where the two values refer the lower/higher frequency measurements, respectively).
    
    A flow of \phc, not exceeding a pressure of 2\,\ub, was pumped through the cell and evacuated by a turbomolecular pump.
    Above this pressure, significant pressure broadening of the rotational lines was observed. 
    
\section{Results and discussion}

\subsection{Spectroscopic considerations}

\ce{PhC3N} belongs to the $C_{2\mathrm{v}}$ symmetry group and possesses a $\Tilde{\mathrm{X}}\,^1\mathrm{A}_1$ electronic ground state with 39 modes of vibration following the irreducible representation $\Gamma = 14 \hspace{1mm} \mathrm{A}_1 \oplus 3\hspace{1mm} \mathrm{A}_2 \oplus 9\hspace{1mm} \mathrm{B}_1 \oplus 13\hspace{1mm} \mathrm{B}_2$.
All vibrational modes except those of $\mathrm{A}_2$ symmetry are IR active; $\mathrm{A}_1$ and $\mathrm{B}_2$ modes correspond to in-plane vibrations (respectively $a$- and $b$-type bands) while $\mathrm{B}_1$ modes are out-of-plane vibrations ($c$-type bands). 
\ce{PhC3N} is a prolate asymmetric top molecule with a permanent dipole moment of 5.9\,Debye along the $a$ axis (axis of the \ce{C3N} bonds, Fig. \ref{fig:phc3n}), according to our calculations.
\ce{PhC3N} also has nitrogen nuclear quadrupole hyperfine structure, but this splitting is only partially resolved at low frequency in the CE-FTMW measurements. Given that the molecule is $C_{2\mathrm{v}}$ symmetry, we also include statistical weights for equivalent exchangeable nuclei for correct transition intensities. Identical to \ce{PhCN}, there are two sets of equivalent hydrogen atoms ($I_1 = I_2 = \frac{1}{2}$) that give rise to Fermi-Dirac statistics for symmetric (even $K$) and antisymmetric (odd $K$) rotational states with a ratio of $10:6$ \cite{mcguire_detection_2018}.

\subsection{Vibrational spectroscopy}

    \begin{figure*}[ht]
        \centering
        \includegraphics[width=\textwidth]{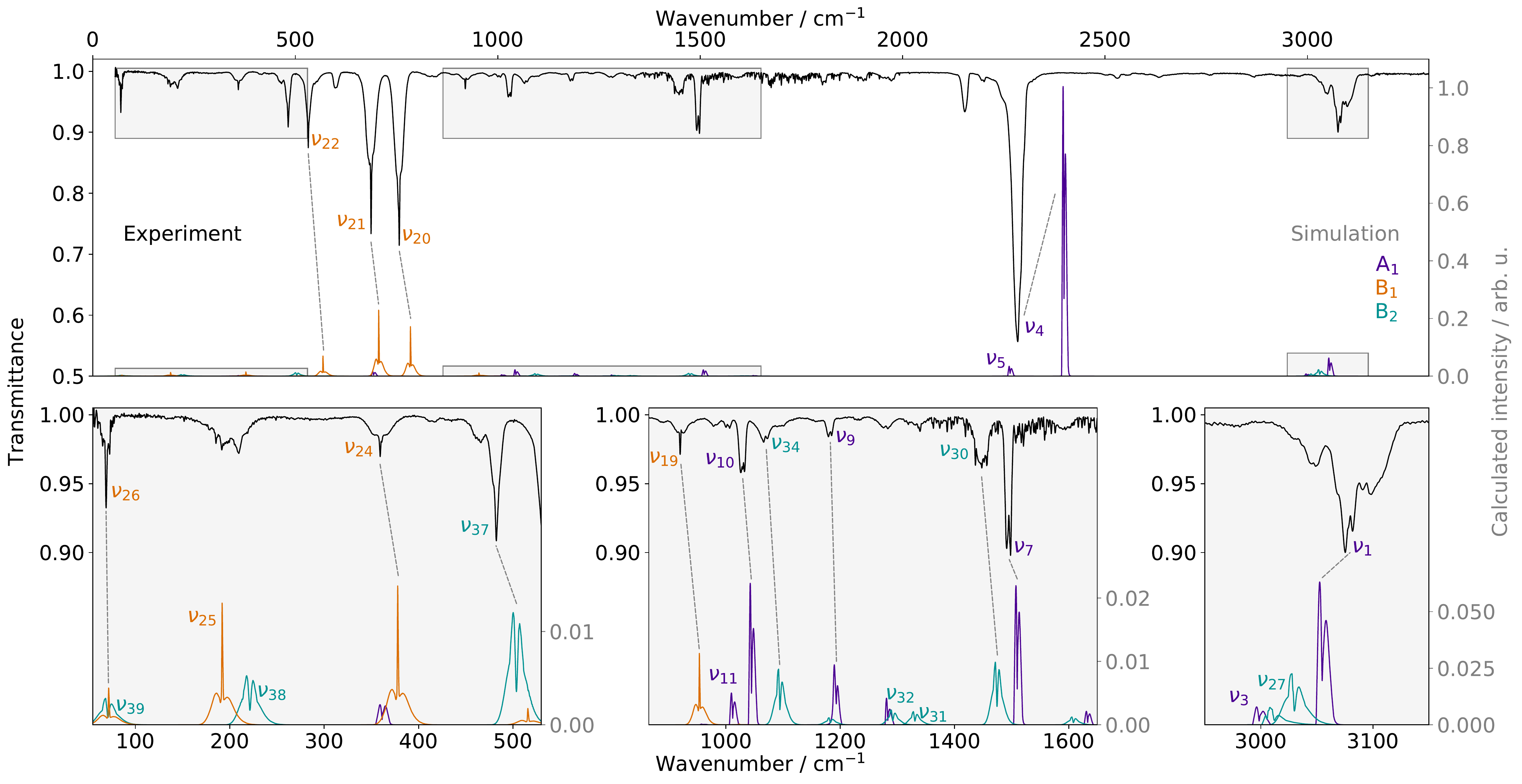}
        \caption{Experimental (top traces, in black) and simulated (bottom traces, at 300\,K, where different colors correspond to different symmetries) vibrational spectrum of \phc. The three bottom panels are expanded portions of the full spectrum, as indicated by gray rectangles.
        Simulation performed using the PGOPHER software \cite{Pgopher} and the results of the anharmonic calculations (band centers, intensities, and rotational constants) and normalized to the strongest IR active mode ($\nu_4$).   The simulations are inverted relative to the observed spectrum solely for comparison purposes. Secure band assignments are indicated by dash lines; additional labelled bands are those for which an assignment is proposed; simulated bands without any labels remain unassigned.}
        \label{fig:IR}
    \end{figure*}

The infrared spectrum of \phc is presented in Fig.~\ref{fig:IR} together with a simulation of the vibrational fundamentals predicted by our anharmonic quantum chemical calculations. 
At our experimental resolution, the rotational contour of most bands is evident in the spectrum. 
Because the simulation is in very good agreement with the experiment, numerous assignments can be made with confidence. 
For ambiguous assignments, two criteria can be invoked: (i) $c$-type bands (B$_1$ symmetry) exhibit a sharp Q-branch; and (ii) $a$-type bands (A$_1$) are usually narrower than the others (as can be seen on the simulated spectrum).
Out of the 36 active infrared modes of \phc, 14 can be assigned with little or no ambiguity while 9 other bands have tentative assignments at this juncture; the remaining bands are either weak or are predicted in crowded regions of the spectrum (see Table \ref{tab:vibAssign} for a detailed list of the proposed assignments). 
Experimental band centers are taken as the frequency of the $Q$-branch when one exists ($c$-type bands), equidistant between the $P$- and $R$-branches, or at the maximum of the envelope when no clear contour is visible. 
Considering $c$-type $Q$-branches are several wavenumbers wide, an accuracy of $\pm 2$\,\wn can be expected for the  band centers with B$_1$ symmetry. 
For the others, a conservative value of $\pm 5$\,\wn is proposed.

\begin{table*}[p!]
    \centering
    \caption{Fundamental vibrational bands (position and intensity) of \phc from the quantum chemical calculations performed in this work at the harmonic and anharmonic levels of theory, and proposition of assignments. Modes are numbered following the anharmonic calculations frequency order. Modes energy are in wavenumbers, intensities are in km/mol, $\delta$ values are in \%. The experimental assignments are split in two categories, the relatively \rev{secure} ones (column ``Assign.") and the tentative ones (column ``Prop."); $\delta$ values of the latter are reported in italics.
    }
    \small
    \label{tab:vibAssign}
    \begin{tabular}{rr c rr c rr c rrr}
    \toprule
\multicolumn{2}{c}{Mode} && \multicolumn{2}{c}{Harm. Calc.} && \multicolumn{2}{c}{Anharm. Calc.} && \multicolumn{3}{c}{Exp.}\\ \cmidrule{1-2} \cmidrule{4-5} \cmidrule{7-8} \cmidrule{10-12}
$\nu$& Sym. && Energy & Int. && Energy & Int. && Assign. & Prop. &  $\delta\,^a$ \\
\midrule
1  & A$_1$ && 3214 & 6.7  && 3055 & 13.5 && 3078 && 0.8 \\				
2  & A$_1$ && 3226 & 5.6  && 3026 & 0.0  && & 3047 & \textit{0.7}			\\	
3  & A$_1$ && 3197 & 0.0  && 2999 & 1.7  && & 3047 & \textit{1.6} \\		
4  & A$_1$ && 2431 & 242.3&& 2399 & 220.1&& 2284	&& -4.8 \\				
5  & A$_1$ && 2292 & 10.7 && 2265 & 7.5  && & 2154/2198 & \textit{-4.9/-3.0} \\
6  & A$_1$ && 1675 & 1.0  && 1633 & 0.5  && 	\\				
7  & A$_1$ && 1545 & 7.7  && 1510 & 4.9  && 1495 && -1.0	\\			
8  & A$_1$ && 1306 & 1.6  && 1284 & 0.8  && 	\\			
9  & A$_1$ && 1214 & 2.1  && 1193 & 2.0  && 1182 &&-0.9	\\			
10 & A$_1$ && 1064 & 4.6  && 1045 & 4.5  && 1029 &&-1.5	\\			
11 & A$_1$ && 1027 & 1.3  && 1012 & 1.0  && & 1004/981 & \textit{-0.8/-3.1}\\
12 & A$_1$ && 981  & 0.1  && 960  & 0.0  && \\					
13 & A$_1$ && 706  & 3.4  && 693  & 3.5  && \\					
14 & A$_1$ && 371  & 0.8  && 362  & 0.5  && \\	\midrule				
15 & A$_2$ && 1019 & 0.0  && 1000 & 0.0  && \\					
16 & A$_2$ && 876  & 0.0  && 862  & 0.0  && \\					
17 & A$_2$ && 413  & 0.0  && 412  & 0.0  && \\		\midrule				
18 & B$_1$ && 1042 & 0.0  && 1023 & 0.0  && \\					
19 & B$_1$ && 967  & 2.8  && 953  & 2.4  && 920&& -3.5 \\				
20 & B$_1$ && 792  & 38.7 && 784  & 32.7 && 757&& -3.4 \\				
21 & B$_1$ && 716  & 36.7 && 706  & 42.7 && 687&& -2.6 \\				
22 & B$_1$ && 576  & 13.2 && 568  & 12.4 && 532&& -6.3 \\				
23 & B$_1$ && 523  & 0.2  && 515  & 0.3  && 	\\				
24 & B$_1$ && 384  & 2.8  && 377  & 2.8  && 359 && -4.9	\\			
25 & B$_1$ && 195  & 2.6  && 192  & 2.5  && &185  & \textit{-3.4} \\		
26 & B$_1$ && 74   & 0.6  && 71   & 0.7  && 69 && -3.3 \\					\midrule
27 & B$_2$ && 3206 & 1.9  && 3030 & 9.5  && &3095  &\textit{2.1}\\		
28 & B$_2$ && 3222 & 8.4  && 3012 & 2.6  && 	\\				
29 & B$_2$ && 1646 & 0.7  && 1608 & 0.5  && 	\\				
30 & B$_2$ && 1493 & 8.6  && 1475 & 4.5  && 1448 && -1.8	\\			
31 & B$_2$ && 1364 & 1.2  && 1331 & 0.9  && &1336 & \textit{0.4} \\	
32 & B$_2$ && 1320 & 1.3  && 1292 & 1.0  && &1280 & \textit{-0.9}\\		
33 & B$_2$ && 1193 & 0.4  && 1183 & 0.5  &&  \\					
34 & B$_2$ && 1117 & 4.4  && 1094 & 3.8  && 1069 && -2.3	\\			
35 & B$_2$ && 644  & 0.0  && 635  & 0.0  && \\					
36 & B$_2$ && 584  & 0.5  && 574  & 0.4  && \\					
37 & B$_2$ && 511  & 5.3  && 504  & 5.6  && 482 && -4.3 \\			
38 & B$_2$ && 226  & 2.3  && 221  & 2.4  && &209  & \textit{-5.6}\\		
39 & B$_2$ && 71   & 1.1  && 71   & 1.2  && &69   & \textit{-2.7}\\	
\bottomrule
    \end{tabular}
    
    \medskip
    \begin{minipage}{0.8\textwidth}
    $^a\ \delta = (\mathrm{Exp.}-\mathrm{Anharm. Calc.})/\mathrm{Anharm. Calc.}\times 100$ 
    \end{minipage}
\end{table*}

A few of the assignments proposed in Fig.~\ref{fig:IR} and Table \ref{tab:vibAssign} warrant further discussion.
While the assignment of the $\nu_{26}$ band (out-of-plane backbone motion) is relatively secure thanks to the presence of a sharp $Q$-branch, the band appears stronger than expected from our calculations. 
A possible explanation is that  $\nu_{39}$ (\ce{C#C-C#N} rocking), predicted roughly at the same frequency as $\nu_{26}$, contributes to the experimental band profile, thus enhancing absorption. 
In absence of any specific feature arising from $\nu_{39}$ on the experimental spectrum, we have chosen to assign this band to the same band center as $\nu_{26}$, at 69\,\wn. 
Slightly higher in frequency are $\nu_{25}$ and $\nu_{38}$,  which involve similar nuclear motion as $\nu_{26}$ and $\nu_{39}$ and are also predicted to lie close in energy (at 192 and 221\,\wn, respectively), which in combination likely yield the feature observed around 200\,\wn. 
It is not straightforward, however, to unambiguously assign each band center, in part because $\nu_{25}$ is the only band of $B_1$ symmetry with a sizeable transition moment that does not exhibit a sharp Q-branch. 
Tentative assignments are proposed in Table \ref{tab:vibAssign}, but their accuracy should be taken with caution as the actual bands centers could lie in a 10 to 20\,\wn window from the proposed assignments. 
Concerning $\nu_{11}$ (ring symmetric breathing) and $\nu_{5}$ (\ce{C#C-C#N} stretch)---predicted at 1012 and 2265\,\wn, respectively---two bands lie systematically close to the expected energy, with reasonable band profiles, thus in each case both assignments are reported in Table \ref{tab:vibAssign}. 
In the case of $\nu_{5}$, the lowest frequency assignment, $i.e.$, 2154\,\wn appears most likely as the relatively strong intensity of the observed band would indicate a fundamental that could otherwise not be predicted. 
However the shape of the band lying at 2198\,\wn is closer to that expected for an $a$-type band, assuming that our aforementioned criterion remains valid at these frequencies (i.e.~hot bands and combination bands could significantly affect the simplistic fundamentals-only picture).
%
    
The most difficult analysis lies in the ${\sim}3000-3200$\,\wn region, where \ce{C-H} stretching motions are observed. From previous studies on similar molecules like phenylacetylene \cite{king_ethynylbenzene_1970} and naphthalene \cite{pirali_high-resolution_2009}, it is well-known that this frequency range is plagued by anharmonic resonances which can significantly complicate assignment. For \ce{PhC3N}, three modes of $\mathrm{A}_1$ ($\nu_1$, $\nu_2$, $\nu_3$) and two of $\mathrm{B_2}$ ($\nu_{27}$, $\nu_{28}$) symmetries (Table \ref{tab:vibAssign}) very close in energy (of order of ${\sim}$10\,\wn for the harmonic frequencies) and can mix strongly, thus qualitatively shifting the fundamental energies and altering band intensities. In such cases,  an approximate deperturbation analysis was performed using the generalized VPT2 calculations to identify and treat anharmonic resonances, e.g.~those arising from Fermi (so-called ``1--2'' resonances) and Darling-Dennison (``1--1'' interactions in the present case). These resonances are identified based on small differences in the energies of states and a model variational calculation \cite{martin_anharmonic_1995}; the former is a zeroth order estimate for resonances, while the latter tests for the magnitude of the coupling \cite{bloino_anharmonic_2015}.

The VPT2 routines in Gaussian treat the problem of state-to-state coupling as effective $2\times2$ Hamiltonians, where the diagonal elements correspond to the state energies, and the off-diagonal elements represent coupling between the two states. These values are shown in Table \ref{tab:resonance}, and give rise to significant deviation from the harmonic frequencies and intensities shown in Table \ref{tab:vibAssign}. Of particular  note is the unintuitive complete loss of intensity in $\nu_2$: $\nu_3$ is ``dark'' in the harmonic approximation, and it gains intensity primarily through borrowing intensity from the stronger $\nu_1$ and $\nu_2$ bands. In a general resonance picture, this interaction shares intensity, and to render a mode completely inactive is extremely rare if not unheard of. Given the VPT2 treatment here is only approximate,  our effective deperturbation analysis is likely inadequate to properly treat these bands, and  a fully coupled model involving fundamentals, combination bands, and overtones, is required instead. 

In light of this preliminary analysis, however, as well as the fact that the ${\sim}3000$\,\wn region is heavily congested, we assume all three $\mathrm{A}_1$ modes are IR active. Measurement under cold conditions---either in a supersonic jet or buffer gas cell---will help  future analysis of this molecule by eliminating the possibility of combination bands and overtones, in addition to minimizing lineshape blending from rotational contours. Similarly, selective deuteration might clarify the assignment of some features \cite{king_ethynylbenzene_1970}.

\begin{table}[ht!]
    \centering
    \caption{Strong anharmonic resonances and their corresponding off-diagonal matrix elements for bands in the ${\sim}3100-3200$\,\wn\,region. Darling-Dennison (DDR) and Fermi (FR) resonances are indicated; in this table, the former corresponds to $1-1$ type DDR, referring to the number of quanta for states involved.} 
    \small
    \begin{tabular}{l l c r}
        \toprule
        State 1 & State 2 & Type & Coupling  \\
        \midrule

       $v_1=1$ & $v_3=1$ & DD & -6.6 \\
       $v_2=1$ & $v_3=1$ & DD & 14.5 \\
       $v_3=1$ & $v_6 = 1 + v_7 = 1$ & FR & -23.2 \\
       $v_3=1$ & $v_{29} =1 + v_{30} = 1$ & FR & -29.3 \\
       $v_{27}=1$ & $v_{28} = 1$ & DD & 9.4 \\

        \bottomrule
    \end{tabular}
    \label{tab:resonance}
\end{table}
    
\subsection{Rotational spectroscopy}

Using the ground state rotational and quartic centrifugal distortion constants from the anharmonic calculation, 65 strong lines of the CP-FTMW spectrum were assigned in a straightforward fashion using the PGOPHER software \cite{Pgopher} (Fig.~\ref{fig:CPMW}).
The derived values are extremely close to those predicted by the calculation (the weighted frequency difference $\delta$ is less than 1\,\% for the rotational constants, see Table \ref{tab:rotconstants}).
This initial set of constants was then used to assign the millimeter-wave data.
Loomis-Wood diagrams were produced by means of the LWWa software from \citet{LODYGA2007} to aid in the assignment of the high-$J$ transitions.
In total, 6151 $a$-type transitions (3780 different frequencies as a result of unresolved asymmetric splitting) of \phc in its ground vibrational state were assigned  in the millimeter-wave spectrum, with values of $J^{\prime\prime}$ up to 199 and $K_a^{\prime\prime}$ up to 42.

The SPFIT/SPCAT suite of programs \citep{Pickett1991} using the Watson $S$-reduced Hamiltonian in the $I^r$ representation was used to determine best-fit spectroscopic constants. 
All transitions were weighted accorded to their expected experimental accuracy, $i.e.$, {2}\,kHz and 25\,kHz for the CE-FTMW and CP-FTMW transitions, respectively, and 50\,kHz for the millimeter-wave transitions.
To reproduce the data to their experimental accuracy, inclusion of several sextic and octic centrifugal distortion constants was required.
Finally, the CE-FTMW transitions---the only ones for which the nuclear quadrupole splitting was resolved---were added to the fit to determine the $\chi$(N) terms. 
All 57 hyperfine components were reproduced  to the measurement uncertainty by adjusting only $\chi_{aa}$(N) and $\chi_{bb}$(N). 
Both parameters are close to those expected from calculation (to within about 15\,\%, Table \ref{tab:rotconstants}).
When a transition was observed by both CP-FTMW and CE-FTMW spectroscopy, only the latter was retained in the fit, owing to the higher frequency accuracy of the cavity instrument.
The rotational constants derived from a fit to all of the assigned rotational transitions are reported in Table \ref{tab:rotconstants}.
The  final {49}\,kHz RMS value of the fit, corresponding to a reduced standard deviation {$\sigma=1.00$}, indicates that our present model adequately reproduces the ground state rotational spectrum of \phc.

\begin{table}[ht!]
    \centering
    \caption{Spectroscopic constants (rotational, centrifugal distorsion, and nuclear quadrupole constants) of \phc in its vibrational ground state (in MHz) and relevant fit parameters. Numbers in parenthesis are $1\sigma$ uncertainties expressed in the unit of the last digit. Parameters in brackets were kept fixed to the calculated values.}
    \label{tab:rotconstants}
    \small
    \begin{tabular}{ld{4.3}d{4.10}r}
    \toprule
Constant                & \multicolumn{1}{c}{Calc.$^a$} & \multicolumn{1}{c}{Exp.} & \multicolumn{1}{c}{$\delta\,^b$}\\
\midrule
$A_0$                   & 5656.9     &   5659.722\,(15) & 0.05   \\
$B_0$                   &  567.0     & 569.582206\,(39) & 0.46   \\  
$C_0$                   &  515.4     & 517.404488\,(37) & 0.39   \\  
$D_J\times 10^{6}$      & 3.6        &    3.85110\,(77) & 7.0      \\
$D_{JK}\times 10^{3}$   & 0.78       &   0.827177\,(85) & 6.1      \\
$D_K\times 10^{3}$      & 0.42       &      0.400\,(68) & -4.8    \\
$d_1\times 10^{6}$      & -0.50       &  -0.55191\,(24) & 10   \\
$d_2\times 10^{6}$      & -0.52      &   -0.57101\,(79) & 9.8     \\
$H_J\times 10^{12}$     &            &     -0.957\,(11) &  \\
$H_{JK}\times 10^{9}$   &            &     0.7916\,(18) &  \\
$H_{KJ}\times 10^{6}$   &            &   -0.03051\,(10) &  \\
$h_2\times 10^{12}$     &            &      0.575\,(15) &  \\
$h_3\times 10^{12}$     &            &    0.10329\,(55) &  \\
$L_{JJK}\times 10^{15}$ &            &     -1.295\,(20) &  \\
$L_{JK}\times 10^{12}$  &            &    0.06954\,(86) &  \\
$L_{KKJ}\times 10^{12}$ &            &     -3.512\,(36) &  \\
\\    
$\chi_{aa}$             &-4.96       &     -4.219\,(77) & -15 \\ 
$\chi_{bb}$             & 2.39       &      2.114\,(68) & -12 \\
$\chi_{cc}$             & 2.57       &      2.143\,^c   & -17 \\
\\
$N\,^d$ & &\multicolumn{1}{r}{6256/3877/57} \\
$J''_\mathrm{max}$, $K''_{a\,\mathrm{max}}$ & & \multicolumn{1}{r}{199, 42}\\
rms /kHz &&  \multicolumn{1}{r}{0.049} \\
$\sigma\,^e$ &&  \multicolumn{1}{r}{1.00}\\
\bottomrule
    \end{tabular}
    
\medskip
\begin{minipage}{\columnwidth} \scriptsize
$^a$ $\omega$B97XD/cc-pVQZ level of theory, Bayesian corrected for $A$, $B$, and $C$, and anharmonic values for the centrifugal distortion constants, and equilibrium values for the hyperfine constants\\
$^b \delta = (B_\mathrm{exp.}-B_\mathrm{calc.})/B_\mathrm{calc.} \times 100$ (in \%)\\
$^c$ Derived value\\
$^d$ Total number of lines in the fit / Number of different frequencies / Number of lines with resolved nuclear quadrupole structure\\
$^e$ Reduced standard deviation, unitless
\end{minipage}
\end{table}

\begin{figure*}[ht!]
    \centering
    \includegraphics[width=\textwidth]{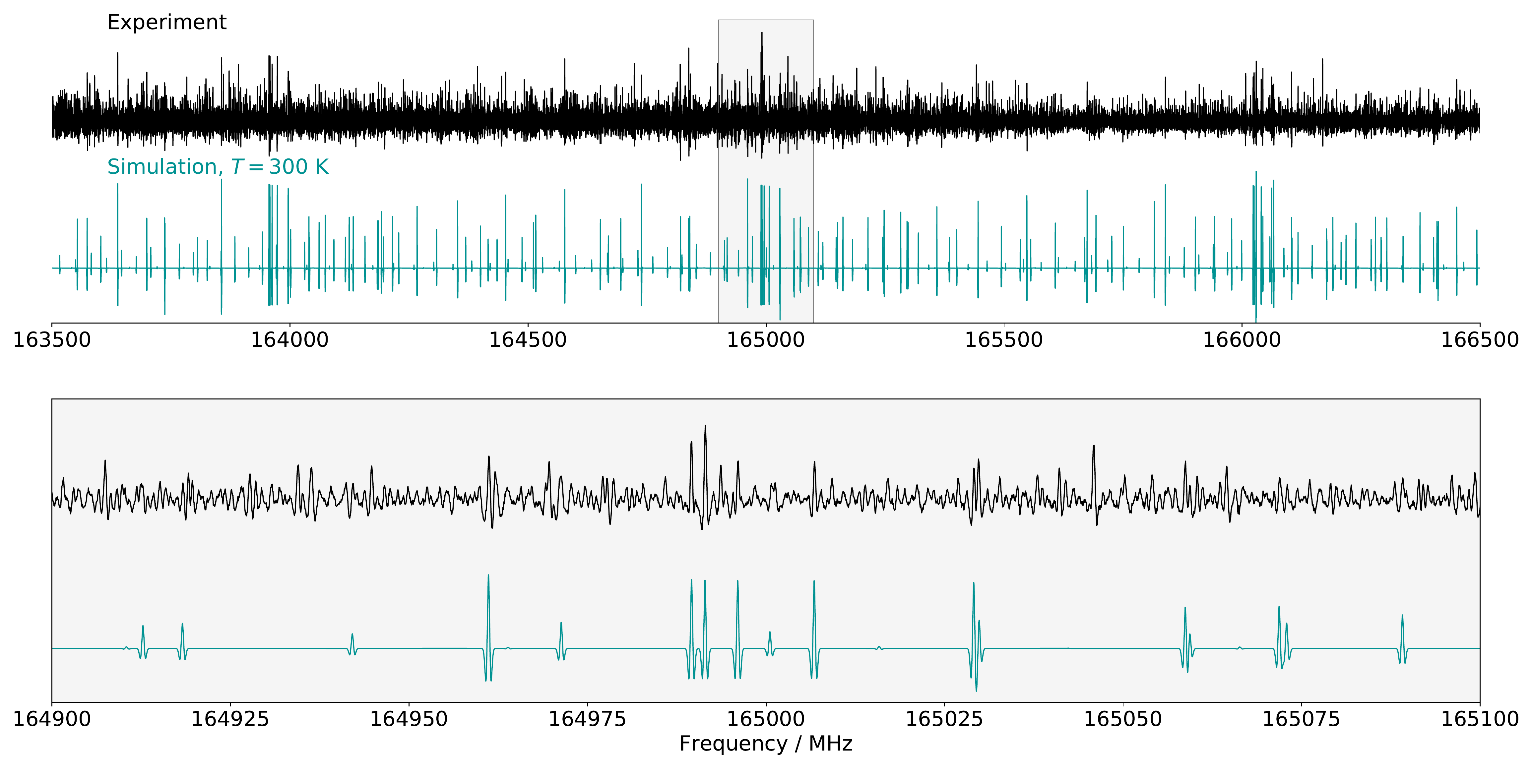}
    \caption{Portions of the millimeter-wave spectrum of \phc in comparison with a simulation of the pure rotational transitions in the ground vibrational state using the best-fit set of spectroscopic constants (Table \ref{tab:rotconstants}). The simulation has been performed using the PGOPHER software and the resulting trace was then post-processed with a second derivative to allow a more straightforward comparison with the experimental spectrum. The line density in the experimental trace is far greater than our simulation, very likely because of lines from vibrational satellites. }
    \label{fig:mmwav}
\end{figure*}

Figure~\ref{fig:mmwav} shows a portion of the millimeter-wave spectrum in comparison with a simulation of \phc in its vibrational ground state using the experimentally determined best-fit  parameters from Table \ref{tab:rotconstants}. 
As illustrated in this figure, many lines remain unassigned, but most of these likely arise from vibrational satellites, for which no attempt at assigning was made in the present study. 
Although longer integration times would allow a more in-depth analysis of these satellites, the spectrum is already fairly dense, implying that we may be close to the confusion limit.
Indeed, the 2\,\textmu bar pressure used in this study, although quite low, was actually a compromise between reasonable signal-to-noise ratio and pressure broadening. 
Even at this pressure many lines are broader then expected from the effects of pressure broadening alone, and consequently may in fact be a spectral superposition of several transitions.

\begin{figure}[ht!]
    \centering
    \includegraphics[width=\columnwidth]{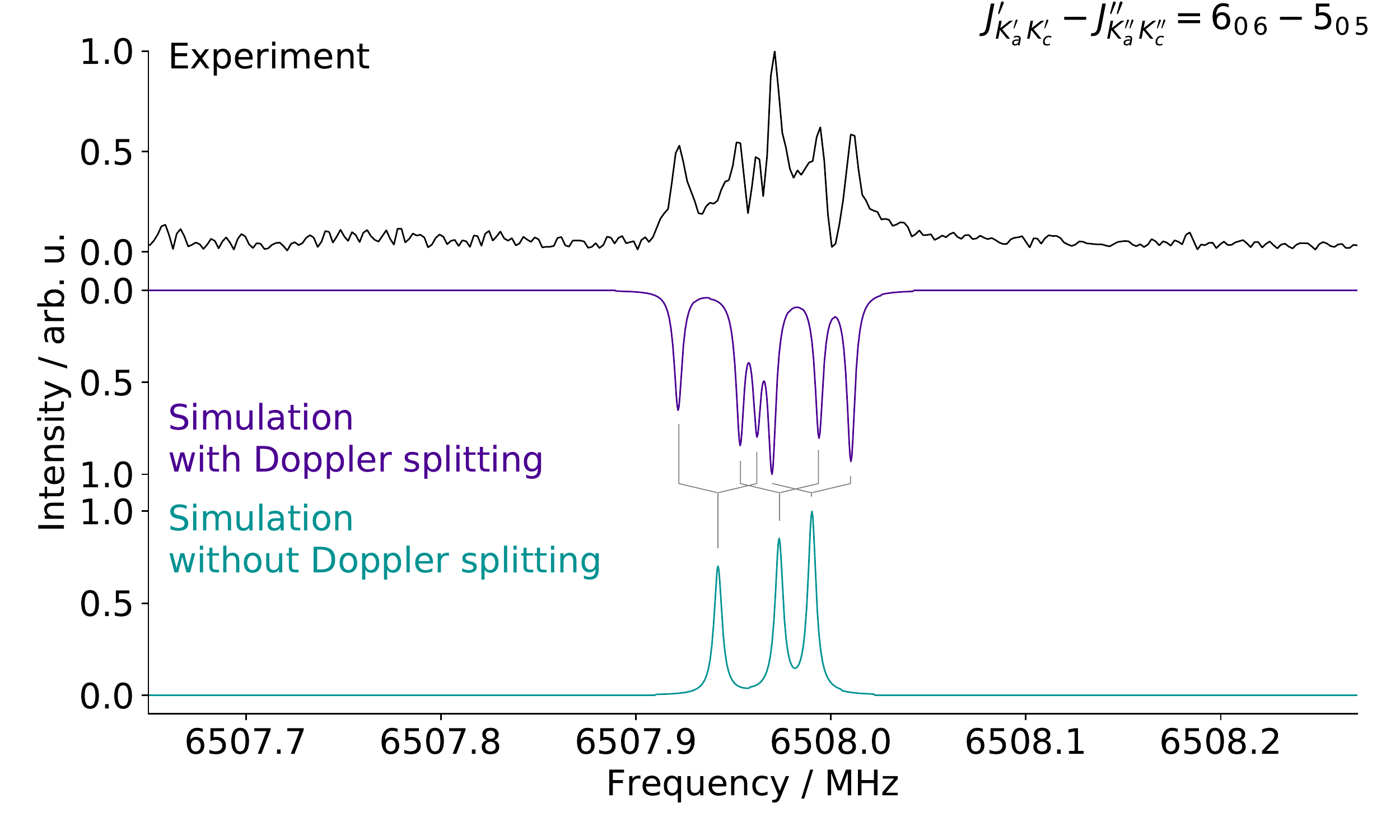}
    \caption{Example of a transition showing resolved hyperfine structure on the CE-FTMW spectrum, and comparison with 10\,K simulations using the final set of spectroscopic parameters, with and without taking into account the Doppler splitting (simulations performed using the PGOPHER software, assuming a Lorentzian lineshape).  As before, the simulation with Doppler splitting is inverted relative to the observed spectrum solely for comparison purposes.}
    \label{fig:cmwavCV}
\end{figure}

Regarding hyperfine splitting, 
its magnitude rapidly collapses with increasing $J$, as expected from this type of interaction. 
No splitting is observed in the CP-FTMW nor in millimeter-wave measurements, and in the cavity experiments it is marginally resolved above $J^{\prime\prime}\approx11$ (${\sim}12$\,GHz). 
Even at the lowest-$J$ transitions measured here ($J^{\prime\prime}=5$, ${\sim}6$\,GHz), the splitting due to hyperfine and Doppler effects are comparable (Fig.~\ref{fig:cmwavCV}).
The experimentally derived value of $\chi_{aa}$(N) [$-4.219$\,(77)\,MHz; Table~\ref{tab:rotconstants}] is very similar to that reported for \ce{PhCN} [-4.23738(36)\,MHz \cite{wohlfart_precise_2008}], and the relative magnitudes are in agreement with the theoretical values of $\chi_{aa}$(N) calculated at the $\omega$B97X-D/cc-pVQZ level of theory for \ce{PhC3N} (-4.959\,MHz) and \ce{PhCN} (-4.962\,MHz). 
The small changes in $\chi$(N) {are} an indication that the local electronic structure of the nitrogen atom is relatively insensitive to the distance from the aromatic ring. 
Equivalently, this finding implies that electron delocalization through ring conjugation is very poorly coupled to the chain regardless of length. Similar behavior is seen for cyanopolyyne chains, where the value of $eQq$ equivalent to $\chi_{aa}$(N) (around $-4.1$\,MHz) is relatively invariant with the length of the chain as well \cite{mccarthy_microwave_2000}.
    
\subsection{Astronomical considerations}

Detection of molecules in space by radio astronomy is heavily dependent on the magnitude of their permanent dipole moment. 
In comparison with \ce{PhCN},  
our theoretical predictions suggest \ce{PhC3N} is substantially more polar (5.9\,D vs.~of 4.5\,D, where the former value has statistical uncertainty of $\pm0.25$\,D based on our prior benchmarking at the $\omega$B97X-D/cc-pVQZ level of theory \cite{lee_bayesian_2020}). 
Currently, \ce{PhCN} is theorized to form in cold, dark clouds via a barrierless reaction between \ce{C6H6} and CN radical \cite{mcguire_detection_2018,lee_gas-phase_2019}.  We thus speculate that \phc could be formed via \ce{C2} insertion to \ce{PhCN} or through an analogous \ce{C#N} addition reaction between \ce{CN} radical and phenylacetylene (\ce{PhC2H}), or \ce{C3N} and \ce{C6H6}. If the latter mode is operative, then the abundance ratio of \ce{PhCN}/\ce{PhC3N} will be dependent on \ce{CN}/\ce{C3N}, assuming similar reaction cross-sections. The \ce{CN + PhC2H} process has been studied in crossed-beam experiments by \citet{bennett_chemical_2010}, where the authors identify \ce{PhC3N} as a potential reaction product, albeit not definitively so due to the lack of isomer specificity and at collision energies well in excess of interstellar cloud conditions (${\sim}30$\,kJ/mol). 
This finding suggests  other isomers, namely \textit{ortho, meta, para}-CEB, might plausibly be formed from this reaction. 
While we have not attempted to experimentally characterized these species, estimates of their spectroscopic parameters are provided here.
By scaling the experimental parameters for \ce{PhC3N} (Table \ref{tab:rotconstants}) to correct for vibrational and electronic effects (Table S6 in the \esi), in conjunction to Bayesian uncertainties obtained from benchmarking \cite{lee_bayesian_2020}, reliable constraints of these constants should aid future experimental searches for  these species.

\begin{figure}[ht!]
     \centering
     \includegraphics[width=\columnwidth]{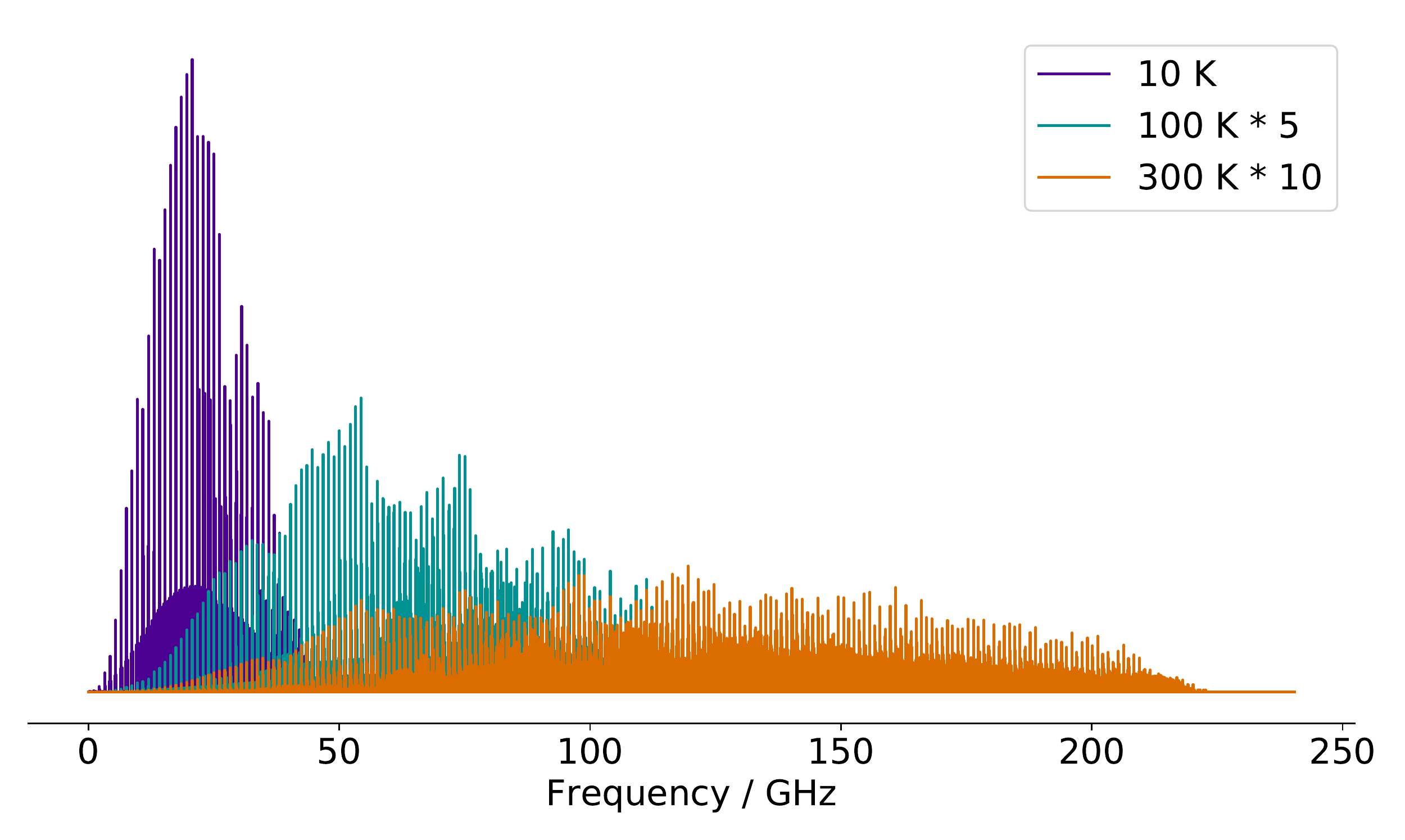}
     \caption{Calculated rotational spectrum of \phc at 10\,K (purple), 100\,K (blue), and 300\,K (orange). For the purposes of display, these relative intensities for the 100\,K plot have increased by a factor of 5, while for the 300\,K plot, the increase is 10.}
     \label{fig:tempDistrib}
 \end{figure}

In terms of astronomical detection, although the hyperfine structure does not take a large part the present rotational analysis, this splitting is partially resolved with cavity measurements up to 16\,GHz. 
In cold, dark clouds such as TMC-1 where the linewidths are comparable to those measured with our CE-FTMW spectrometer, it is thus necessary to consider hyperfine splitting, as recent work on \ce{PhCN} \cite{mcguire_detection_2018} demonstrates. 
This is particularly true for a relatively heavy molecule like \phc whose strongest lines should lie at centimeter-wavelengths at low temperatures (Fig.~\ref{fig:tempDistrib}): at 10\,K, the strongest features correspond to $J^{\prime\prime}=21$ around 23\,GHz, while at 6\,K ---a typical temperature for molecules in TMC-1--- the strongest features are near 15\,GHz ($J^{\prime\prime}=14$). 
Thus, the $\mathrm{X/K_u}$ (8--12/12--18 GHz) bands appear to be the most promising to detect \phc in TMC--1. 
In sources with somewhat warmer temperatures, the intensity of individual transitions is significantly decreased due to the larger partition function, and  the peak intensity, although relatively flat,  falls in the W (75--110 GHz) and N (100--200 GHz) bands. 

\begin{figure}
    \centering
    \includegraphics[width=0.48\textwidth]{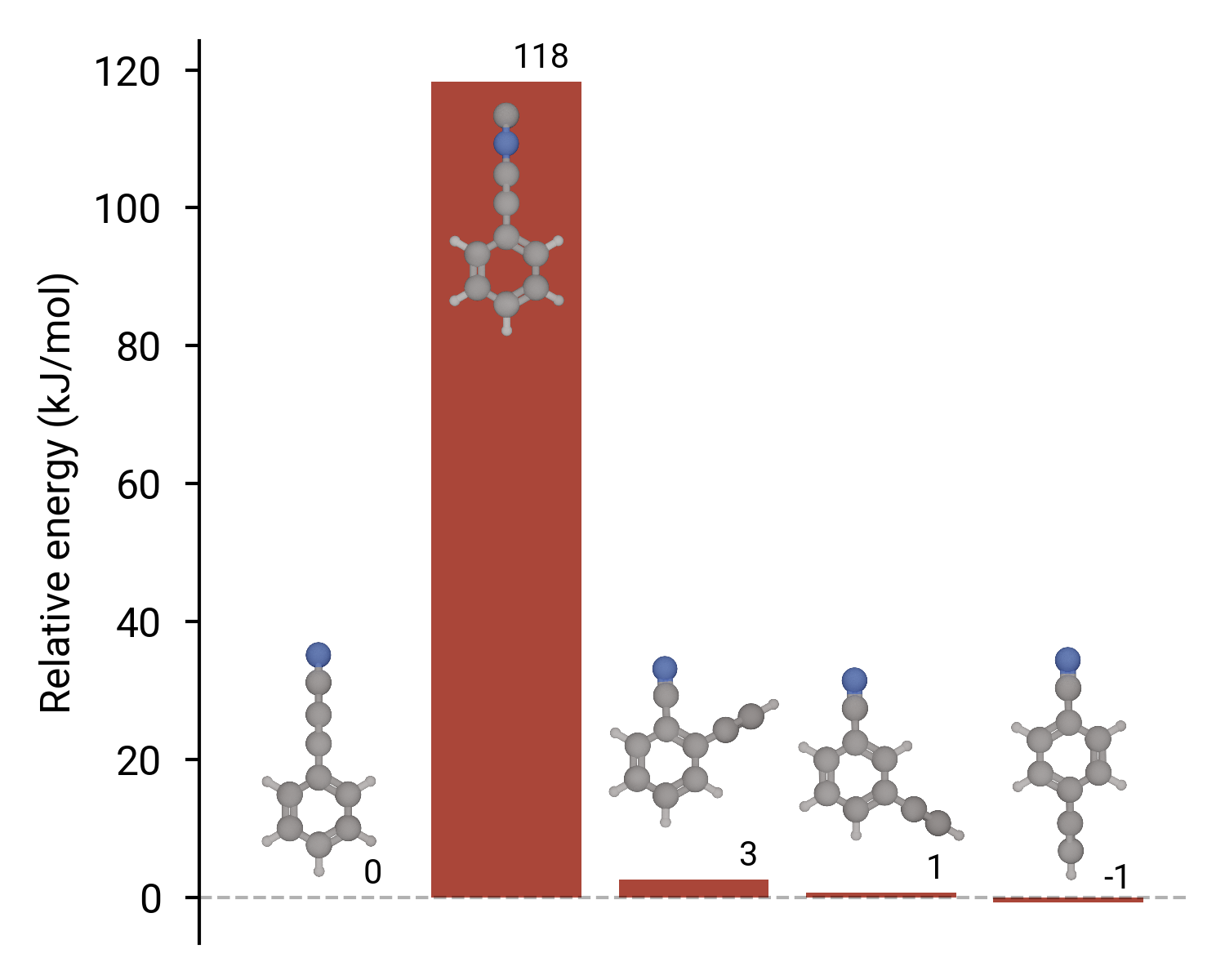}
    \caption{0\,K energetics of isomers of interest calculated at the G3//B3LYP level of theory, relative to \ce{PhC3N}. Energies to the nearest kJ/mol are annotated above each bar.}
    \label{fig:pes}
\end{figure}

Detection of multiple isomers is also a powerful tool for constraining physical and chemical conditions in astrophysical environments. To assist in this process, we have calculated the relative energy of the aforementioned \textit{ortho, meta,} and \textit{para}-CEB isomers, along with the isocyanide isomer, \ce{PhCCNC}. 
As shown in Fig.~\ref{fig:pes}, apart from the isocyanide isomer, placement of the acetylenic unit on different parts of the ring produces isomers with comparable stability to \ce{PhC3N}, i.e.,~they are effectively degenerate at the level of uncertainty afforded by G3//B3LYP ($\pm4$\,kJ/mol). 
As such, a determination of their relative abundances would provide a sensitive test of thermodynamics vs.~kinetics in molecule formation. 
To aid further laboratory and hopefully astronomical efforts, Table S6, in addition to providing estimates of rotational constants, also reports calculated dipole moments at the $\omega$B97X-D/cc-pVQZ level of theory using two methods of empirical scaling to correct for vibrational effects and deficiencies in the electronic structure method. 
We note that the Bayesian scaling factors obtained in Ref.~\citep{lee_bayesian_2020} applied to \ce{PhC3N} ---where we now have accurately determined parameters--- exceed the performance of the purely theoretical VPT2 corrections, and bring the theoretical predictions within a few MHz of the experimentally measured ones (see also Fig. S1). 
We expect a similar degree of precision and uncertainty for other isomers.

\section{Conclusion}

Using a combination of gas-phase measurements and quantum chemical calculations, the fundamental vibrational frequencies of most of the strongest IR-active fundamentals of \phc and over 6000 pure rotational transitions in its ground state between 6.5 and 220\,GHz have been measured. 
The  assignment of spectra in different regions was guided by theoretical predictions: in the infrared, anharmonic calculations helped to assign most bands, with the exception of the most congested and perturbed region around $3000$\,\wn; in the radio domain, estimates of centrifugal distortion and hyperfine terms proved useful. Comparisons with \ce{PhCN} suggest that the local electronic structure of the terminal nitrogen ---as probed through its quadrupole moment $\chi_{aa}$(N)--- is relatively unaffected by chain lengthening, indicating a similar electric field gradient for nitrogen in both \ce{PhCN} and \ce{PhC3N}. In addition, we provide accurate and reliable predictions of the thermochemistry and spectroscopic parameters for several isomers of \phc, which should prove useful in guiding future laboratory experiments. These isomers include the isocyanide isomer of \phc and the three cyanoethynylbenzenes alluded to in previous work \cite{bennett_chemical_2010}; their discovery along with \phc in the ISM would provide sensitive measurements of local chemical and physical interstellar environments.

With highly precise measurements of the rest frequencies and corresponding spectroscopic constants, a search for \phc can now be undertaken with considerable confidence in the ISM. The large permanent dipole moment (predicted to be 5.9\,D), in addition to its chemical and structural similarity to astronomical \ce{PhCN} makes \phc an excellent candidate for detection towards cold, dark molecular clouds such as TMC-1.  Hyperfine-resolved measurements are expected to be highly relevant in a potential discovery, given that at low temperatures the strongest transitions lie in the $\mathrm{X/K_u}$ bands where the splitting is comparable to the source linewidth.

\section*{Acknowledgements}
O.C., O.P., and M.-A.M.-D. acknowledge funding support from the R\'egion Ile-de-France through DIM-ACAV$^+$, from the \textit{Agence Nationale de la Recherche} (ANR-19-CE30-0017-01), from the ``\textit{Investissements d'Avenir}'' LabEx PALM (ANR-10-LABX-0039-PALM), and from the Programme National ``Physique et Chimie du Milieu Interstellaire'' (PCMI) of CNRS/INSU with INC/INP co-funded by CEA and CNES.
K.L.K.L. and M.C.M. acknowledge funding support from NSF grant AST-1908576 and NASA grant 80NSSC18K0396.
Z.S.B. acknowledges support from the Chateaubriand Fellowship of the Office for Science \& Technology of the Embassy of France in the United States. 
The authors are thankful to the AILES beamline staff for providing access to the FTIR interferometer. 

\bibliographystyle{model1-num-names}
\bibliography{biblio.bib}

\end{document}